\documentstyle[aas2pp4]{article}

\begin{document}

\lefthead{Makino}
\righthead{Post-Collapse evolution}

\title{Post-Collapse evolution of globular clusters}

\author{Junichiro Makino}

\affil{Department of Graphics and Information Science, \\
College of Arts and Sciences, University of Tokyo\\
3-8-1 Komaba, Meguro-ku, Tokyo 153, Japan}

\def\versiondate{{9 Nov. 1995}}
\def\titleabbrev{{Post-Collapse evolution}}


\begin{abstract}
A number of globular clusters appear to have undergone core collapse,
in the sense that their predicted collapse time is much shorter than
their current age. Simulations using gas models and Fokker-Planck
approximation have shown that the central density of a globular
cluster after the collapse undergoes nonlinear oscillation with large
amplitude (gravothermal oscillation). However, whether such an
oscillation actually takes place in a real $N$-body system has
remained unsolved, because an $N$-body simulation with a sufficiently
high resolution would have required the computing resource of the
order of several Gflops$\cdot$years. In the present paper, we report
the result of such a simulation, performed on a dedicated
special-purpose computer GRAPE-4. We simulated the evolution of
isolated point-mass systems with up to 32,768 particles. The largest
number of particles reported previously is 10,000. We confirmed that
gravothermal oscillation takes place in an $N$-body system.  The
expansion phase shows all signatures that are considered as the
evidences of the gravothermal nature of the oscillation. At the
maximum expansion, the core radius is $\sim 1$\% of the half-mass
radius for the run with 32,768 particles. The maximum core
size $r_c$ depends on $N$, as $<r_c> \propto N^{-1/3}$.

\end{abstract}

\def\etal{{\it et al.~}}

\newpage

\section{Introduction}

A number of globular clusters appear to have undergone core collapse
(\cite{Cohn1984}, \cite{Hut1992}). Djorgovski and King
(1986) showed that 15\% of galactic globular clusters have unresolved
density cusps. Recent observations with HST have demonstrated that
some of these clusters have the core sizes smaller than 0.03 arcsec,
which is as far as one can go with present technique ($e.g.$, \cite{Sosin1995}, \cite{Yanny1994}).

The theoretical study of the evolution after the core collapse was
pioneered by Henon (1975), who incorporated the energy production by
binaries into his Monte-Carlo calculation. He found that the whole
cluster expanded homologously in the thermal timescale. Similar
results were obtained by gaseous models and Fokker-Planck
calculations.

Sugimoto and Bettwieser (1983, Bettwieser and Sugimoto 1984) found
that the post-collapse expansion is unstable if the energy production
is inefficient. In the unstable case, the central density showed an
oscillation with a large amplitude. They called this oscillation
gravothermal oscillation.  They modelled the post-collapse evolution
of globular clusters using a conducting gas sphere with artificial
energy production. The efficiency of the energy production is related
to the total number of particles $N$, and for large $N$ efficiency is
small. Thus, their result implies an $N$-body system with number of
particles larger than a critical number should show the gravothermal
oscillation.

Since several other calculations based on similar models did not
reproduce the result of Sugimoto and Bettwieser (1983), the validity
of their result had been controversial for a few years after their
first paper.  However, calculations with improved accuracy confirmed
that the oscillation takes place in both the gaseous models and
Fokker-Planck (FP) calculations(\cite{Goodman1987}, \cite{Heggie1989},
\cite{Cohn1989}).

What have been found in gas models and FP calculations are summarized
as follows. There is a homologously expanding solution for the
post-collapse evolution of the cluster (\cite{Goodman1984}). In this
solution, the size of the core is proportional to $N^{-1/3}$. Thus,
the core is smaller for larger number of particles. If the core is too
small, this homologous expansion becomes unstable in a way similar to
the way in which an isothermal sphere is unstable, and the core starts
to show an oscillating behavior. The critical number of particles is
around 7,000 (Goodman 1987). If the total number of particle is
slightly larger than this critical value, the oscillation is
regular. However, at $N\simeq 10,000$, ``period doubling'' takes place
and the density shows oscillation with double peaks. For $N > 50,000$,
oscillation becomes apparently chaotic (\cite{Heggie1989},
\cite{Breeden1994}).

Both the contraction and the expansion are driven by the gravothermal
instability, and the energy production by binaries acts as the trigger
to start the expansion. During the expansion, the thermal energy flows
in from the halo to the core, and this heat flux is supported by
so-called ``temperature inversion''. This temperature inversion itself
is formed through the expansion of the core driven by the binary
heating in the following way.

Binaries deposit the energy to the core through ``indirect heating''
(\cite{Hut1985}). When a binary hardens through interaction with a third
star, that star is likely to be kicked out of the core, because the
recoil velocity is much larger than the velocity dispersion of the
core. As a result, the binding energy of the core becomes smaller. The
core expands to reestablish the dynamical equilibrium. This expansion
causes the velocity dispersion of the core to decrease. Thus the
central temperature becomes lower than the temperature of the region
just outside the core.

Once the central temperature becomes lower than the temperature around 
the core, the heat starts to flow inward. This inward heat flux let
the core expand further, and the temperature of the core continues to
decrease.  This self-supported expansion can continue as far as the
temperature gradient outside the core is small.  In practice, however, 
temperature gradient becomes larger as the core size becomes larger,
and the temperature inversion vanishes at a certain core size. After
that, the core starts to recollapse. 

The critical number of particles for multi-component systems is larger
than that for single-component systems, and for a ``realistic'' mass
function, the critical number of particle can be as large as $3 \times
10^5$ (\cite{Murphy1990}).

The question whether such gravothermal oscillation would takes place
in real globular clusters or in $N$-body systems has not been settled
yet. Both gaseous models and FP models have many simplifying
assumptions that might make the evolution completely different from
that of a real cluster. For example, both assume spherical symmetry,
while in $N$-body simulations cores are known to wander around
(\cite{Makino1987}, \cite{Heggie1994}). Both assume that the energy
production by binaries is smooth, while in $N$-body system binaries
are formed stochastically and emit energy intermittently.

An $N$-body simulation with a sufficiently large number of particles
has been impossible, simply because the requirement for computer power
has been excessive. Makino \etal (1986) performed 100-body simulations
of an equal-mass cluster, using both softened and unsoftened
potentials. They used a softened potential in order to reduce the
energy production from binaries.  They found oscillatory behavior for
both cases, but with larger amplitude for softened potential. They
argued that the fact that the amplitude of oscillation is larger for
reduced energy production implies that the observed oscillation is
driven by the gravothermal instability and not purely by the heating
by binary.  McMillan and Lightman (1984, \cite{McMillan1986a})
developed hybrid calculation code in which the central region of a
cluster is treated as an $N$-body system while the outer region is
treated in FP formalism. They observed oscillatory behavior but
concluded that it was not gravothermal. Inagaki (1986) performed 1000-
and 3000-body simulations but found no sign of oscillation. He used
standard Salpeter mass function as the distribution of the masses of
particles. Makino (1989) performed simulation of a 3000-body
equal-mass system and saw some signs of oscillation, including the
temperature inversion. Spurzem and Aarseth (1996) performed 10,000
body simulation and saw oscillatory behavior, but they did not see any
clear evidence of the gravothermal oscillation.  $N$-body simulations
in the last decade can be summarized as follows: The central density
showed some oscillatory behavior, and its amplitude is seemingly
larger for larger number of particles. However, each expansion phase
is associated with large energy input from one or a few binaries. As a
result, whether the expansion is driven by instability or by energy
input has been unclear. In addition, in all but one simulations, the
number of particles was smaller than the critical value of 7,000.  The
10,000 body simulation by Spurzem and Aarseth (1996) was not long
enough to draw clear conclusion. Therefore it was not clear whether
oscillatory behavior is caused by gravothermal instability or not.

Heggie (1989, see also \cite{Heggie1994}) tried to see whether
gravothermal expansion would take place if one constructs an $N$-body
system that has an initial temperature inversion. They constructed an
$N$-body system which mimics the density and temperature structures of
the expansion phase of the gas model calculation by Heggie and
Ramamani (1989), and follow the evolution of the $N$-body system. They
found that the core expands in the thermal timescale. Thus, they at
least proved that the core of an $N$-body system can expand
gravothermally.

In the present paper, we describe the result of $N$-body simulations
with number of particles 2,048--32,768. This is the first calculation
with the number of particles well beyond the critical number of 7,000
which covers sufficiently long time after the collapse to determine
the nature of the post-collapse evolution. All simulations were
performed on GRAPE-4 (\cite{Taiji1996}), a special-purpose computer
for collisional $N$-body simulations. Our main results are the
following. First, for large $N$, the core density and core mass
exhibited oscillations with large amplitude.  The core mass at the
maximum expansion is in good agreement with FP or gas model results
(1\% of the total mass for 16k- and 32k-particle runs).  Second, we
confirmed that the observed oscillation is driven by gravothermal
instability.  Several long expanding periods without any energy input
were observed. The temperature inversion was visible in such expansion
phases.  The behavior of the core density is strikingly similar to the
result of FP calculations with stochastic heat source
(\cite{Takahashi1991}), which strongly suggests that the mechanism is
the same. The trajectory in the central density-central temperature
plane clearly indicates that the inward heat flux supports the
expansion.

The structure of this paper is the following. In section 2, we
describe the initial model, numerical method and the computer used. In
section 3 we present the result. Section 4 is for discussion.

\section{Model and Numerical method}

\subsection{Initial models and the system of units}

We followed the evolution of isolated systems of point-mass particles. 
For all calculations, we used random realizations of the Plummer model
as the initial condition. The system of units we adopted is the
standard (Heggie) units (\cite{Heggie1986}), in which $G=1$, $M=1$,
and $E=-1/4$, where $G$ is the gravitational constant, $M$ is the
total mass of the cluster, and $E$ is the total energy of the cluster. 
All particles have the same mass $m=1/N$, where $N$ is the total
number of particles. In this unit, the half-mass crossing time
$t_{hc}$ is $2\sqrt{2}$.

\subsection{Numerical method}

For all calculations, we used NBODY4 (Aarseth 1985, 1996), modified
for GRAPE-4 (Taiji \etal 1996). The numerical integration scheme
adopted in NBODY4 is the 4th order Hermite scheme (\cite{Makino1991a},
\cite{Makino1992}) with the individual (hierarchical) timestep
algorithm (\cite{McMillan1986b}, \cite{Makino1991b}) to use the GRAPE
hardware or parallel/vector computers efficiently. Close two-body
encounters and stable binaries are handled by KS regularization
(\cite{Kustaanheimo1965}). Special treatment for compact
few-body subsystems is also possible.

The 4th-order Hermite scheme has several advantages over traditional
ABM type schemes. It is a self-starting scheme and therefore much
easier to implement than the multistep ABM scheme of the same
order. Its local truncation error has much smaller coefficient, which
allows a larger timestep for the same accuracy. On general-purpose
computers, this advantage is partly cancelled by the additional CPU
time to calculate the first time derivative, but on GRAPE-type
machines this is an significant advantage (\cite{Makino1992},
\cite{Makino1993}).

The output are taken at intervals of a fixed time, which is a certain
fraction of the crossing time. We recorded central density,
core radius, number of particles in the core,  radii of Lagrangian
shells, velocity dispersion within Lagrangian shells, binaries
and their binding energies. The core parameters are calculated
following Casertano and Hut (1985), with modifications described in
McMillan \etal (1990).

The accuracy parameter of the time integration is adjusted so that the
energy error between two outputs is smaller than a certain prescribed
value. The value we used is $1\times 10^{-5}\sim 1\times 10^{-6}$
depending on $N$. We required higher accuracy for larger $N$, since
the duration of the simulation is longer. When the energy error is
very large, the program automatically reads the output at the previous
checkpoint and restarts with a reduced accuracy parameter.

The $N$-body simulation of point-mass systems with a large number of
particles poses several technical problems beside the necessary
computing power. The difficulty comes from the wide range of the
timescale. The critical separation of two particles at which we apply
the KS regularization is about $1/N$. The timestep for particles
involved is of the order of $1/(100N^{1.5})$. On the other hand, the
timescale of the evolution of the system is of the order of $N$. Thus,
the timesteps of particles that are integrated in KS formalism is
$1/100N^{2.5}$ of the system time, which, for $N\sim 10^5$, about as
small as the limit of double precision numbers. Note that this
estimate is valid for a soft binary. A hard binary would require 
timesteps smaller by several orders of magnitude than described above. Of course,
a clever treatment can overcome this kind of difficulties, but the
amount of programming work and the complexity of the resulting code
would be considerable. The simplest solution would be to use the
longer number format ($e.g.$, the quadruple precision). This would
have been a viable solution in '70s, when mainframe machines could
handle quadruple precision arithmetic reasonably fast.  Unfortunately,
most of present RISC computers are extremely slow at quadruple
precision arithmetic.

\subsection{Hardware and calculation cost}

For all calculations, we used GRAPE-4 (\cite{Taiji1996}). The GRAPE-4
is a special-purpose computer designed to accelerate the $N$-body
simulation using the Hermite integrator and hierarchical timestep
algorithm. The total system consists of 1692 pipeline processor chips
and has the theoretical peak speed of 1.08 Tflops. The simulations
reported in the present paper were performed while the assemblage and
testing of the GRAPE-4 system were underway. Thus the number of
processors varies during the calculation. For most of the 32k particle
run, we used one quarter of the system with 423 processor chips, which
has the peak speed of 270 Gflops. Calculations with smaller $N$ were
performed while the available number of processors was smaller.

The actual performance depends on many factors, but most strongly on
the number of particles. With the present host computer (DEC
3000/900), average speed we got for 32k-particle run was about 50
Gflops. This 32k-particle run took about three CPU months. In
comparison, the 10k-particle run reported by Spurzem and Aarseth
(1996) took two CPU months on a Cray YMP. Since the calculation cost
of globular cluster simulation is proportional to $N^{3.3}$, one
quarter of GRAPE-4 is effectively 50 times faster than a Cray YMP.

At present, a 32k-particle system is about the largest system we can
try for equal-mass case.  Since the calculation speed of GRAPE-4 is
still limited by the speed of the host computer, the actual
calculation speed is roughly proportional to $N$ for $ N< 2\times
10^5$. Thus, the CPU time for globular cluster simulation on GRAPE-4
is proportional to $N^{2.3}$.  A 64k-particle run would take about one
year.

The above estimate is for an equal-mass isolated cluster of point-mass
particles. The computational cost of more realistic calculations
depends on many factors. Among them, the mass spectrum and the
presence of primordial binaries are most important, and have adverse
effects.

Systems with a mass spectrum evolves faster than equal-mass systems (see,
$e.g.$, \cite{Inagaki1984}, \cite{Murphy1988}). Therefore it is
possible to handle a larger system. For example, if the evolution of
the system is $p$ times faster, GRAPE-4 can handle the number of
particles $\sqrt{p}$ times larger. Thus, if the collapse is 4 times
faster, GRAPE-4 can finish a 32k-particle run in 3 weeks or
64k-particle run in 3 months.

The presence of the primordial binaries would increase the CPU
time. At present, however, actual cost is difficult to estimate
because the calculation cost depends strongly on the core size, of
which we have rather little knowledge (see section 4). It should be
noted that the number of floating point operations used to follow the
life of a binary depends only weakly on the total number of particles
in the system (\cite{Makino1990}). Therefore, while the CPU time for
the orbit integration of single stars would increase as $N^{2.3}$, the
cost of binaries would increases only as $N$. It is likely that for
large $N$ the cost of binaries is relatively small, even for
simulations with primordial binaries.

\section{Result}

\subsection{The core oscillation}

Figure 1 shows the time evolution of the central density for all runs.
The time is scaled so that thermal timescale is the same for all runs. 
The scaling factor is $t_r(1000)/t_r(N) = 212.75\log(0.11N)/N$ (Giersz
and Heggie 1994). The core density shows oscillations with a large
amplitude in calculations with large $N$ ($>$16k).  No matter what is
the real nature of this oscillation, it is at least clear that the
core density of the $N$-body system shows oscillation with the
amplitude comparable to that observed in gas models or FP
calculations.

\begin{figure}
\plotone{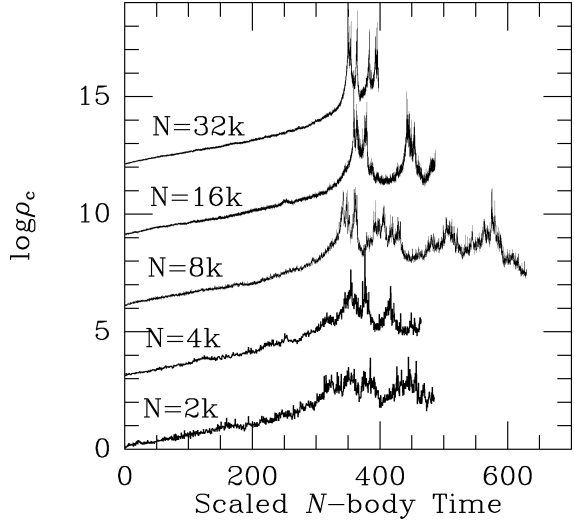}
\caption{The logarithm of the central density plotted as a function of
the scaled $N$-body time. Curves for different values of
$N$ are vertically shifted by 3 units.}\label{fig1}
\end{figure}

For runs with small $N$ (2k and 4k), there are some oscillation-like
features but they are hardly distinguishable from fluctuations. This
result is similar to that of previous studies ($e.g.$ Makino 1989).
For large values of $N$, however, the oscillation with large amplitude
is clearly visible.

Note that in figure 1 there is no clear transition from stable
expansion to oscillation or from regular oscillation to chaotic
oscillation, which were observed in gas and FP models
(\cite{Heggie1989}, \cite{Breeden1994}). The reason is that binaries
emit energy intermittently (\cite{Makino1987}).  Takahashi and Inagaki
(1991) incorporated this stochastic nature of the energy source to
their FP model and found that the core density shows chaotic
oscillatory behavior even if the energy production rate is larger than
the critical value for transition between stable expansion and regular
oscillation. They also found that the amplitude of the oscillation is
smaller for larger energy input (smaller $N$), which is consistent
with the present result. In fact, it would be difficult to distinguish
between our $N$-body calculation results from their stochastic FP
results. Only visible difference is that their result is smoother
while the central density is low. This is because the FP result is
perfectly smooth as far as there is no binary heating.

\subsection{Analysis of the  32k run}

In the previous section, we overview the post-collapse evolution of
$N$-body point-mass systems with 2k-32k particles. In this section, we
take a closer look of the 32k-particle simulation to see whether we
can find direct signature of the gravothermal expansion. It is
generally believed that a long expansion phase without significant
energy input and a temperature inversion during the long expansion
phase are the most direct signatures of the gravothermal expansion
(\cite{Bettwieser1984}, \cite{McMillan1995}). In
this section we investigate both of them.

Figure 2 shows an enlarged view of the time variation of the core
radius as compared with the sum of  binding energies of all
binaries.  It is clear that most of the energy is generated when the
core is very small. 

\begin{figure}
\plotone{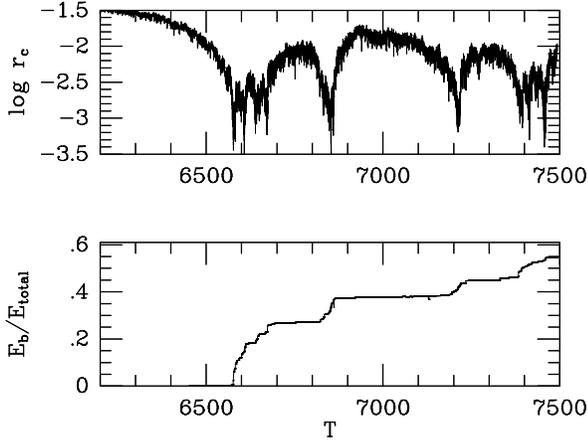}
\caption{The core radius (top) and the binding energy of binaries
(bottom) as a function of time for the 32k-particle run.}
\label{fig2}
\end{figure}
The continued expansion without energy input from binaries has been
considered to be the most direct signature of the gravothermal
oscillation. In figure 2, we clearly see three such expansions, for
$t=6700-6740$, $t=6860-6920$, and $t=7220-7280$.  All these expansions
continue for more than 10 half-mass crossing times, which is hundreds
of the core relaxation time. These expansions cannot be driven simply
by  binary heating. If the expansion were driven only by binary
heating, it could not continue without energy input for the timescale
more than 100 times longer than the core relaxation time.

Figure 3 shows temperature profiles for the contracting and
expanding phases. Near the end of the expanding phase the temperature
inversion of the order of 5\% is clearly visible. Since these
profiles are time-averaged over 10 time units (80 snapshots), the
actual inversion might be somewhat stronger. Note that the temperature
inversion is visible only near the end of long expansion phases also
in gas model and FP calculations (\cite{Bettwieser1984}, 
\cite{Cohn1989}).

\begin{figure}
\plottwo{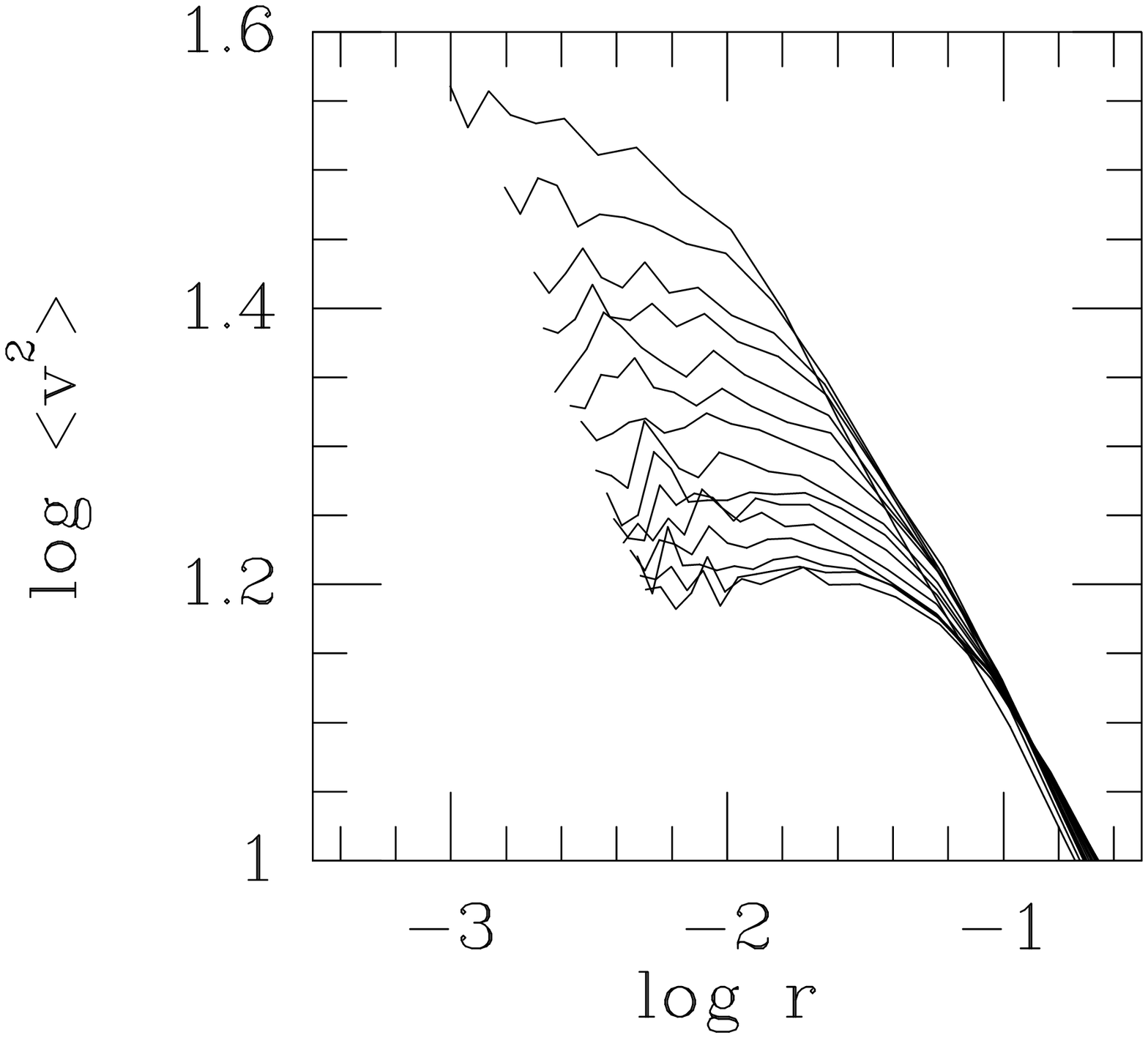}{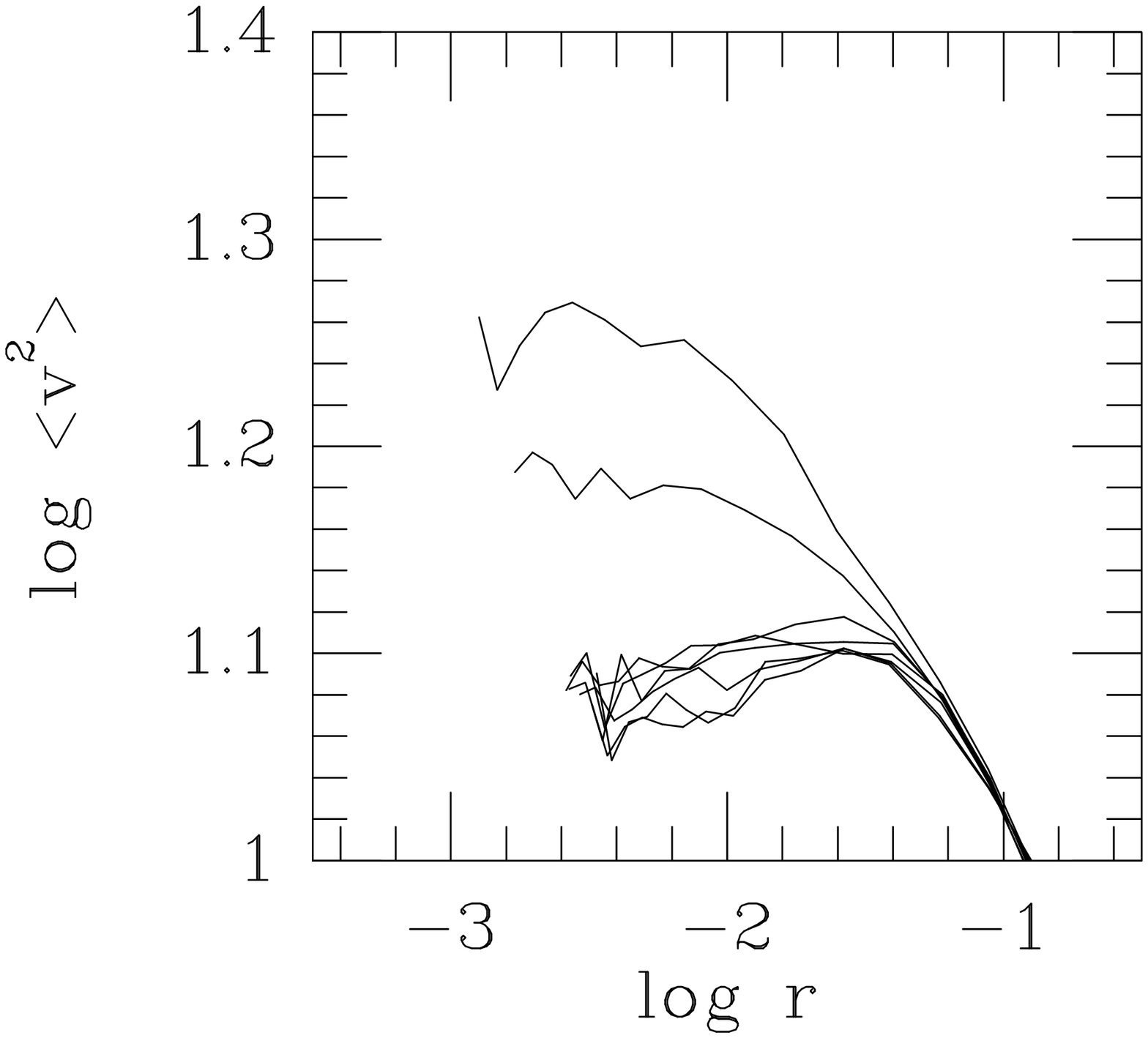}
\caption{Velocity dispersion profiles for (a) contracting and (b)
expanding phases. Each profile is obtained by time averaging over 80
snapshots (10 time units). The time interval between curves is 5 time
units.} \label{fig3}
\end{figure}

Figure 4 shows the relation between the central density and the
central velocity dispersion.  The trajectory shows striking
resemblance to what obtained by gas model calculation
(Goodman 1987).  The fact that the trajectory shows clockwise
rotations means that this is a refrigeration cycle, in which the
central region absorbs heat when the temperature is low, and
release heat when the temperature is high (Bettwieser and Sugimoto
1984, Bettwieser 1985). In particular, the later phase of the large
expansions (indicated by the arrow marked ``B'') is nearly isothermal.
Since the binding energy of binaries is unchanged during this phase,
this nearly isothermal expansion is driven by the heat supplied from
outside the core.  In other words, the expansion is gravothermal.

\begin{figure}
\plotone{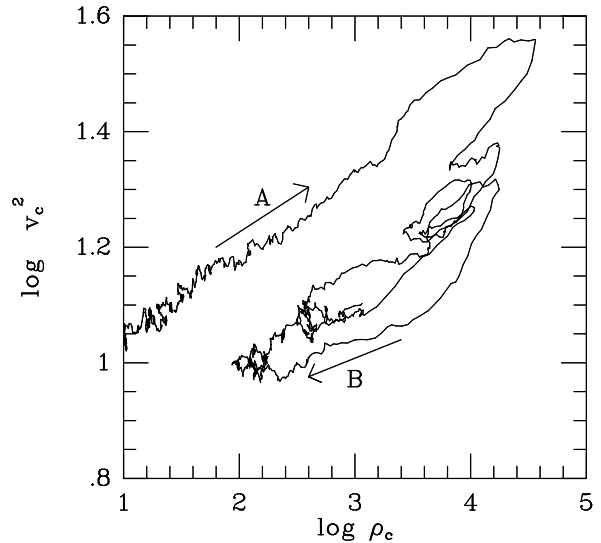}
\caption{The change of the central density and the central velocity
dispersion. Each data points is time-averaged value over 80
snapshots. Arrows indicate the direction of evolution.}
\label{fig4}
\end{figure}

\subsection{The core size --- result and interpretation}

Figure 5 shows the evolution of the number of particles in the
core. For all runs, the number of particles at the maximum contraction
is of the order of 10, while that at the maximum expansion is $\sim 1\%$
of the total number of particles. This result is again in good
agreement with gas models and FP calculations ($e.g.$, \cite{Heggie1989}
\cite{Breeden1994}). 

\begin{figure}
\plotone{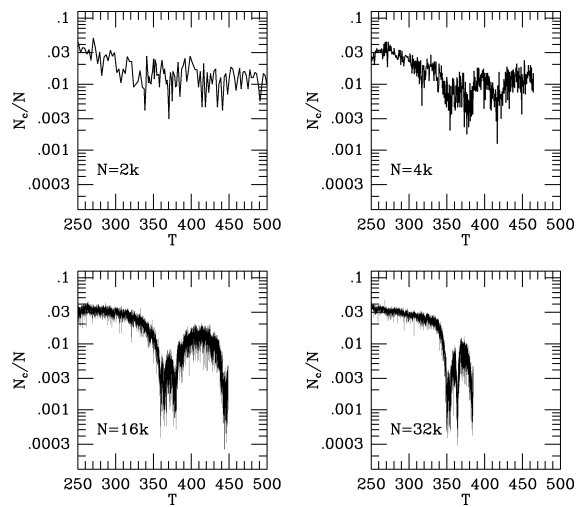}
\caption{The number of particles in the core as the function of the
scaled time, for simulations with 2k, 4k, 16k, and 32k particles}
\label{fig5}
\end{figure}

Figure 6 shows the fraction of time for which the number of particles
in the core is smaller than the value $N_c$ as a function of $N_c/N$,
for the post-collapse phase. In other words, this figure shows the
cumulative distribution of the core mass.  For $N > 8192$, the median
core mass is around 0.5-0.6 \%. This corresponds to $r_c/r_h \sim
0.01$. For $N=32,768$, the core mass is somewhat smaller than that for
16k or 8k runs. 

\begin{figure}
\plotone{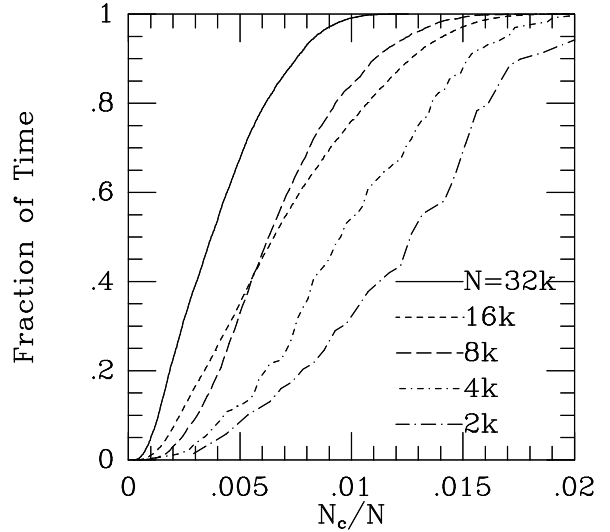}
\caption{Fraction of time for which the number of particles in the
core is smaller than $N_c$, as a function of $N_c/N$.}
\label{fig6}
\end{figure}

Whether the core size in the gravothermal oscillation phase depends on
the total number of particles or not is not well understood. Bettwieser
and Sugimoto (1984) argued that clusters spend most of the time in the
most expanded state, and that the structure of the cluster at the
maximum expansion does not depend on the total number of
particles. However, it is clear from figure 6 that the cumulative time
is roughly proportional to the core radius, and that the coefficient
is larger for larger $N$. In other words, the size of the core depends
on $N$.

Simple theoretical argument suggests that the core size must depend on
$N$. The time-averaged energy production in the core would be the same as
that for the stable expansion (\cite{Goodman1987}, \cite{Heggie1989}). 
This requirement pose a constraint on the time-averaged core size. The
energy production rate is expressed as
\begin{equation}
{dE  \over dt} \propto M_c \rho_c^2,
\end{equation}
where $M_c$ and $\rho_c$ are the core mass and core density. Here we
ignored the dependence on the velocity dispersion, since the inner
part is almost isothermal. For $M_c$ and and the core radius, $r_c$,
we have the relation  $M_c \propto r_c$, again because the inner part
of the cluster is almost isothermal. Thus we have
\begin{equation}
{dE  \over dt} \propto M_c^{-3}.
\label{eqn:erateonmc}
\end{equation}
The time-averaged energy production rate must be equal to that of the
homologous expansion solution obtained by (Goodman 1984). The core
size for the homologous expansion is $N^{-2/3}$ (Goodman 1984, Giersz
and Heggie 1993). Thus we obtain
\begin{equation}
 <M_c^{-3}>^{-1/3} \propto N^{-2/3},
\label{eq:mcpower}
\end{equation}
where $<x>$ means the time averaging. 

Figure 7 shows the arithmetic mean and the maximum of the core mass
after the collapse as functions of $N$. We also plot the quantity $ M_e
=<M_c^{-3}>^{-1/3}$, since we have the theoretical prediction only for
this quantity.  For maximum core mass, the FP result by Breeden \etal
(1994) are plotted for $N=5\times 10^4$ and $10^6$. These values are
read by eye from their figure 5.  The quantity $M_e$ indeed shows
reasonable agreement with the theoretical prediction of equation
(\ref{eq:mcpower}).

\begin{figure}
\plotone{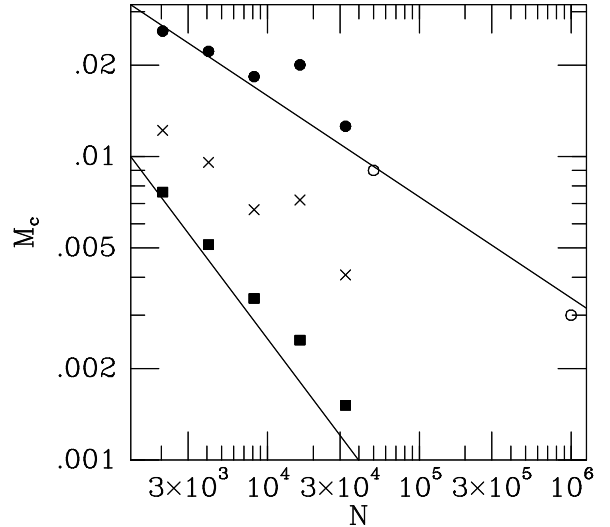}
\caption{Core mass plotted as a function of the total number of
particles $N$. Filled squares are $M_e$ (see text). Filled crosses are 
arithmetic means. Filled circles are maximum values from present
$N$-body simulations. Open circles are maximum values from FP
calculations of Breeden \etal (1994). Straight lines
indicates $M_c\propto N^{-1/3}$  (top) and  $M_c\propto N^{-1/3}$
(bottom), respectively.} 
\label{fig7}
\end{figure}

However, the average and maximum core radii shows
the dependence noticeably weaker than the theoretical prediction. It
should be noted that our $N$-body results for the maximum core mass
and the FP results of Breeden \etal (1994) are on one line expressed
as $M_c \propto N^{-1/3}$.  From figure 7, the core mass at the maximum
expansion is approximated as
\begin{equation}
M_{c,\max} \sim 0.006\left({N \over 10^5}\right)^{-1/3}M.
\end{equation}
The slope of the average core size seems to be somewhat steeper. 

Qualitatively, it is natural that the ratio between the maximum core
mass and $M_e$ is larger for larger $N$, since the ratio between the
minimum core mass and $M_e$ is larger for larger $N$.  The minimum
core mass is $O(1/N)$, since the number of particles in the core at
the core bounce is always around 10-30. In other words, the energy
production rate at the core bounce is larger than the time-averaged
rate by a factor proportional to $N$. Thus, roughly speaking, the core
mass has to stay at the value larger than the value at the core bounce
for most of time. The typical relaxation timescale of the core not at
the maximum contraction is, therefore, $N$ times longer than that at
the maximum contraction. Since the relaxation time is proportional to
$M_c^2$, this implies that the typical core mass is $N^{-1/2}$.  This
is in fact a reasonable fit for the average core mass in figure 7.

One interesting question is
whether it is possible to distinguish the core in the gravothermal
oscillation and the core dominated by primordial binaries. The
theoretical prediction by Goodman and Hut (1989) gives $r_c/r_h \sim
0.02$. The Fokker-Planck calculation by Gao \etal (1991) gave similar
result. On the other hand, $N$-body simulations by McMillan \etal
(1990) gave $r_c/r_h \sim 0.2$.  McMillan \etal (1990) used 1136
particles. If their $N$-body results can be extrapolated to larger
values of $N$, cores with primordial binaries and cores in
gravothermal oscillation would be clearly distinguishable. On the
other hand, FP result by Gao \etal (1991) indicates that the typical
core size at the ``primordial binary burning'' phase and that at the
gravothermal oscillation phase would not be much different.  To obtain
a definitive answer, we need to perform simulations of clusters with
primordial binaries using the number of particles larger than that
employed by McMillan \etal (1990).

\subsection{The core size --- comparison with observations}

Djorgovski and King (1986) showed that 15\% of galactic
globular clusters have unresolved density cusp. They classified these
clusters as ``post core collapse'' (PCC). Recent ground-based
observations (\cite{Lugger1995}) and HST results (\cite{Sosin1995},
\cite{Guhathakurta1994}, \cite{Yanny1994}) demonstrated that some
of the PCC clusters have the cusps unresolved even with HST
(NGC6624, M15). For these clusters, the surface density profile shows
the power law down to 0.3 arcsec. Other clusters (M30), though the slope
seems to levels off toward the center, turned out to be difficult to
determine the slope accurately because there are too few stars. In the
case of M30, the core radius can be anywhere smaller than 1.5 arcsec.

Many researchers claimed that there is a good agreement
between the theory and observation of the core radius. For example,
Djorgovski and Meylan (1994) found that no observed cluster has the
core radius smaller than 1\% of the half-light radius, using the data
compiled by Trager \etal (1993), and argued this to be in good
agreement with the model that assumes the binary-dominated core for
PCC clusters (Vesperini and Chernoff 1994). 

This apparently good agreement between the ``observation'' and
``theory'' is actually due to the fact that both of them overestimated
the core size.  On the observational side, High-resolution
observations have shown that the real core size of the PCC clusters is
significantly smaller than what is assumed in Djorgovski and Meylan
(1994). Trager \etal (1993) adopted the HWHM as the core size of PCC
clusters. Since the HWHM has no relation to the real core radius of
the PCC clusters, it is quite natural that observations with higher
resolution obtained smaller core radii.  On the theoretical side, our
present result suggest that the typical size of the core in the
gravothermal oscillation phase is significantly smaller than the
previous claims.

Our result suggests that  the typical  size of the core in the
gravothermal oscillation phase is smaller than previous estimates
because of the following two reasons. First, the time-averaged core
size is about a half of the maximum size. Second, the core size
depends on $N$ as $N^{-1/\alpha}$, with index alpha $2\sim 3$. 

For NGC6624 and M15, core sizes seem to be too small for a stably
expanding cluster, no matter what is the heat source. So it is highly
likely that they are undergoing the gravothermal oscillation. 

\section{Discussion}

We performed direct $N$-body simulation of the post-collapse evolution
of globular clusters. We confirmed that the gravothermal oscillation
actually takes place in a point-mass  $N$-body system.

Whether real globular clusters undergo gravothermal oscillation or not
is a question which requires further research. If clusters contain
many primordial binaries, even after the core collapse they might
still be burning the primordial binaries. In addition, the effect of
two-body binaries to the evolution of the cluster is still unclear. In
fact, the cross section for the binary formation by tidal capture is
not fully understood yet (Mardling 1995a, 1995b).

The most straightforward way to study the effect of primordial
binaries or two-body capture is the direct $N$-body simulation. In
principle, we can put primordial binaries and their evolution into FP
calculation as Gao \etal (1991) did. However, the standard
one-dimensional FP calculation, which assumes the isotropic
distribution, is not appropriate to follow the binary population,
since most binaries are formed in the core and their orbits are nearly
radial. Thus, we have to solve the FP equation at least in three
dimensions ($E$, $J$ and the binary binding energy $E_b$). It would
require prohibitively large computer power. In addition, reliable
two-dimensional FP calculation has become possible only very recently
(Takahashi 1995). One could also use the Monte-Carlo approach, but its
result must be compared with $N$-body simulation anyway.

We demonstrated that the $N$-body simulation with the number of
particles close to real globular clusters is now possible, thanks to
the extremely powerful special-purpose computer and its full-time
availability. We are now able to use the direct $N$-body simulation to
study various aspects of evolution of globular clusters.

Our 32k-particle calculation took about three months of CPU
time on 1/4 of GRAPE-4. If we tried to do a similar calculation on a
Cray T90 vector supercomputer, it would have taken several years of
CPU time. It is simply impossible to do on present-day supercomputers. 
If we want to finish the calculation in, say, one CPU month, we need a
computer 50-100 times faster than a Cray T90, which will be
available in 10 years from now.

The CPU time of three months is still very long. However, for many
simulations, we do not need as many as 32k particles. A 16k-particle
calculation was finished in 2-3 weeks, on 1/8 of GRAPE-4. Thus to
run many simulations of 16k particle system is now practical.

If we can continue the development of the special-purpose computer, we
will have a system 100-1000 times faster than present GRAPE-4 in the
next five years. Such a system will make it possible to run 50-100k
particle simulation routinely, while 500k-1M particle simulation will
still take months.

I thank Sverre Aarseth, who made the NBODY4 program for GRAPE-4 system
available. I'm grateful to Makoto Taiji and Toshikazu Ebisuzaki, who
developed the GRAPE-4 system in collaboration with myself and Daiichiro
Sugimoto.  I thank Daiichiro Sugimoto for many stimulating discussions
and helpful comments on the manuscript. I also thank Piet Hut, Steve
McMillan, Yoko Funato and Toshiyuki Fukushige for stimulating
discussions.  This work was supported by Grant-in-Aid for Specially
Promoted Research (04102002) of the Ministry of Education, Science,
Sports and Culture, Japan.

\end{document}